\documentclass[10pt,twocolumn]{article}

\usepackage[utf8]{inputenc}
\usepackage[T1]{fontenc}

\usepackage{graphicx}

\usepackage{lipsum}
\usepackage{xcolor}

\usepackage{authblk}
\usepackage[margin=2cm]{geometry}
\usepackage{mathtools, cuted}

\usepackage{bm}
\usepackage{amsfonts}
\usepackage{amsmath}
\usepackage{amssymb}
\usepackage{mathrsfs} 
\usepackage{siunitx}

\title{Measurement of the temperature decrease in evaporating soap films}
\author[1]{François Boulogne}
\author[1]{Frédéric Restagno}
\author[1]{Emmanuelle Rio}
\affil[1]{Université Paris-Saclay, CNRS, Laboratoire de Physique des Solides, 91405, Orsay, France.}

\date{\today}
\begin{document}

\twocolumn[
    \begin{@twocolumnfalse}
        \maketitle
        \begin{abstract}
Recent advances have demonstrated that evaporation can play a significant role on soap film stability, which is a key concern in many industrial areas but also for children playing with bubbles.
Thus, evaporation leads to a film thinning but also to a film cooling, which has been overlooked for soapy objects.
Here, we study the temperature variation of an evaporating soap film for different values of relative humidity and glycerol concentrations.
We evidence that the temperature of soap films can decrease after their creation up to 8$~^\text{o}$C. 
We propose a model describing the temperature drop of soap films after their formation that is in quantitative agreement with our experiments. 
We emphasize that this cooling effect is significant and must be carefully considered in future studies on the dynamics of soap films.
        \end{abstract}
    \end{@twocolumnfalse}
]

%
%

The stability of soap bubbles and films has implications in many unsuspected areas.
The aerosols created during the bubble bursting are involved in the exchange of liquid between the ocean and the atmosphere \cite{Boucher2013,Murphy1998,Leeuw2011,Veron2015}, but also in air pollution, for example above swimming pools, and in toxin aerosolization \cite{Baylor1977,Blanchard1989,Netz2020a} as well as in the expel of flavors on top of carbonated drinks \cite{Liger-Belair2009}.
Induced aerosolization is also relevant in the glass industry, where they are at the origin of major defects \cite{Bolore2018} or in geological physics such as gas exsolution in magmatic chambers \cite{Gonnermann2007}.
For surface stabilized films, the bursting is due to film thinning \cite{Poulain2018PRL,Miguet2020a}.
This thinning rate is fixed both by the liquid flow in the film, the so-called capillary or gravity drainage as identified in the historical studies \cite{Mysels1959}, and by the evaporation.
Recently, the significance of evaporation on film thinning and rupture has been demonstrated
\cite{Li2012,Pigeonneau2012,Poulain2018JFM,Poulain2018PRL,Champougny2018,Miguet2020a}.

However, evaporation not only changes the film thickness.
An additional effect is the existence of global cooling due to evaporation.
The theoretical framework describing cooling-induced evaporation has been developed in the early 20th century, in particular by Houghton \cite{Houghton1933} who studied the evaporation of a small spherical drop.
The liquid temperature is the result of an energy balance between heat diffusion and latent heat of evaporation.
This approach successfully describes the cooling effect \cite{Erbil2012,Tran2018} and has been applied for different applications such as meteorology \cite{Beard1971,Andreas1995} and virus transport \cite{Netz2020a}.
Although this effect is often considered in studies devoted to drop evaporation, to the best of our knowledge, the significance of cooling-induced evaporation is not mentioned in the literature on soap films and foams.
To quantify the significance of evaporation-induced cooling for soapy objects, we propose in this Letter to measure and model the temperature variation of an evaporating soap film for different ambient humidity values and glycerol concentrations.

%
%

\begin{figure}[h!]
    \centering
    \includegraphics[width=.6\linewidth]{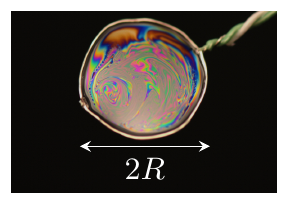}
    \caption{Photograph of a soap film hanging on a frame constituted of a thermocouple probe.
    The radius of the soap film in this picture is $R=6$~mm.}
    \label{fig:setup}
\end{figure}

Soap solutions are made by mixing a dishwashing soap (Fairy with a concentration in surfactant: 5--15 \%) to a mixture of water and glycerol.
We define the glycerol concentration as $\Gamma_{\rm g}(t) = m_{\rm g} / (m_{\rm g} + m_{\rm w}(t))$ where $m$ is a weight and the subscripts $_{\rm g}$ and $_{\rm w}$  stands for  glycerol and water, respectively.
Glycerol is used both to increase the soap film lifetime and to tune the evaporation rate.
We found that at a concentration of $10$~\% and above, the films are sufficiently stable to record the full temperature dynamics.

\begin{figure}
    \centering
    \includegraphics[width=\linewidth]{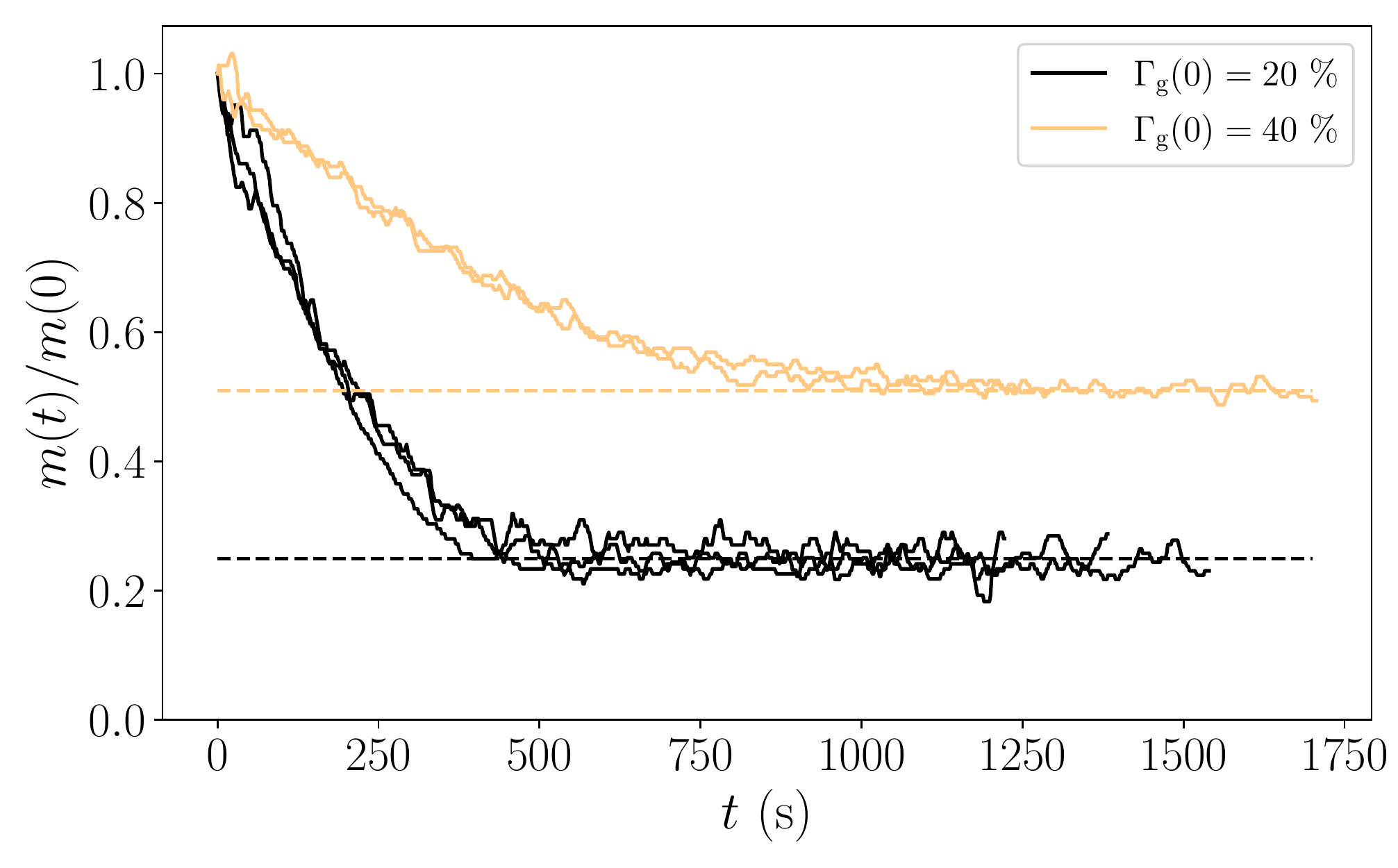}
    \caption{Dynamics of the weight of four soap films normalized by the initial weight.
    The experiments are repeated for two initial glycerol concentrations: 20 and 40~\%.
    The dark dashed line corresponds to $m_{\rm eq}/m(0) = 0.25$ and the light one to $m_{\rm eq}/m(0) = 0.51$.
    The relative humidity is ${\cal R}_{\rm H}=50$~\% and the ambient temperature is about $T_\infty = 21~^\text{o}$C.
    }
    \label{fig:weight}
\end{figure}

To quantify the evaporation dynamics, we record the weight of four identical soap films as shown in figure \ref{fig:setup} on a precision scale (Ohaus Pioneer 210 g) with a precision of 0.1~mg.
The soap films are produced by plunging four circular frames of radius $R=2$ and $R=6$~mm in a soap solution at a temperature $T_\infty$ corresponding to the room temperature.
From the weight measurements, we know that the typical initial film thickness is $h\simeq 1$ $\mu$m.
Experiments are performed in a glove box regulated in humidity \cite{Boulogne2019}.
In figure~\ref{fig:weight}, the weight loss normalized by the initial weight is plotted for two initial glycerol concentrations at a relative humidity of ${\cal R}_{\rm H}=50$~\%.
In both cases, two regimes are observed, composed of a decrease followed by a plateau.
A larger initial concentration of glycerol leads to a slower evaporation and also a larger plateau value.
To interpret these observations, we must consider the vapor pressure of the solutions.

The soap film is placed in an environment of temperature $T_\infty$ and of relative humidity $\mathcal{R}_{\rm H} = p_\infty / p_{\rm sat}^0(T_\infty) $, where $p_\infty$ is the partial pressure in vapor and $p_{\rm sat}^0(T_\infty)$, the saturated vapor pressure of water.
A common phenomenological description of the vapor pressure is provided by Antoine’s equation
\begin{equation} \label{eq:Antoine_equation}
    p_{\rm sat}^0(T) = p^\circ\, 10^{A - \frac{B}{C + T}},
\end{equation}
where $p^\circ = 10^5$~Pa and $A$, $B$, $C$ are constants.
For water at $T \in [0, 30]~^\text{o}$C, $A, B, C$ are obtained by fitting the data extracted from~\cite{Lide2008}, and we obtain $A = 5.34$~K, $B = 1807.52$~K, and $C = -33.90$~K.

The soap film is composed of surfactant, water, and glycerol, the latter being a nonvolatile solvent. 
The vapor pressure depends on the film composition that we consider to be dominated by the water-glycerol ratio.
The saturated pressure of a water-glycerol mixture is

\begin{equation}
    p_{{\rm sat}} (\Gamma_{\rm g}, T) =  p_{{\rm sat}}^0(T)  \dfrac{1 - \Gamma_{\rm g}}{1 + \Gamma_{\rm g} (a-1) },
    \label{eq:psatGly}
\end{equation}
\noindent
where $a = 0.248$ \cite{Glycerine1963}. 
The saturated pressure is, therefore, a decreasing function of the glycerol concentration.

Coming back to the observations made in figure~\ref{fig:weight}, the decrease of the evaporation rate during the dynamics is attributed to the increasing concentration of the nonvolatile solvent, which decreases the saturated pressure of the solution and so, the evaporation rate.

At equilibrium, the temperature of the soap film corresponds to $T_\infty$ and the vapor pressure of the film equals the partial pressure in the atmosphere, \textit{i.e.} $p_{\rm sat}(\Gamma_{\rm g}(t), T_\infty) = p_\infty$.
The equilibrium of pressures lead to the concentration of glycerol $\Gamma_{\rm g}^{\rm eq}$ that reads
\begin{equation}
    \Gamma_{\rm g}^{\rm eq} = \frac{1 - {\cal R}_{\rm H}}{ 1 + (a - 1) {\cal R}_{\rm H}}.
\end{equation}
In our conditions of temperature and humidity, the film reaches an equilibrium at a glycerol concentration $\Gamma_{\rm g}^{\rm eq}\simeq 0.80$.
Therefore, the weight ratio is $m_{\rm eq}/m(0)= \Gamma_{\rm g}(0) / \Gamma_{\rm g}^{\rm eq}$.
For the two glycerol concentrations considered in figure~\ref{fig:weight}, the weight ratios are 0.25 and 0.50, for 20 and 40~\% initial concentrations, respectively.
Thus, the predicted final state is in excellent agreement with the experiments.

\begin{figure}
    \centering
    \includegraphics[width=\linewidth]{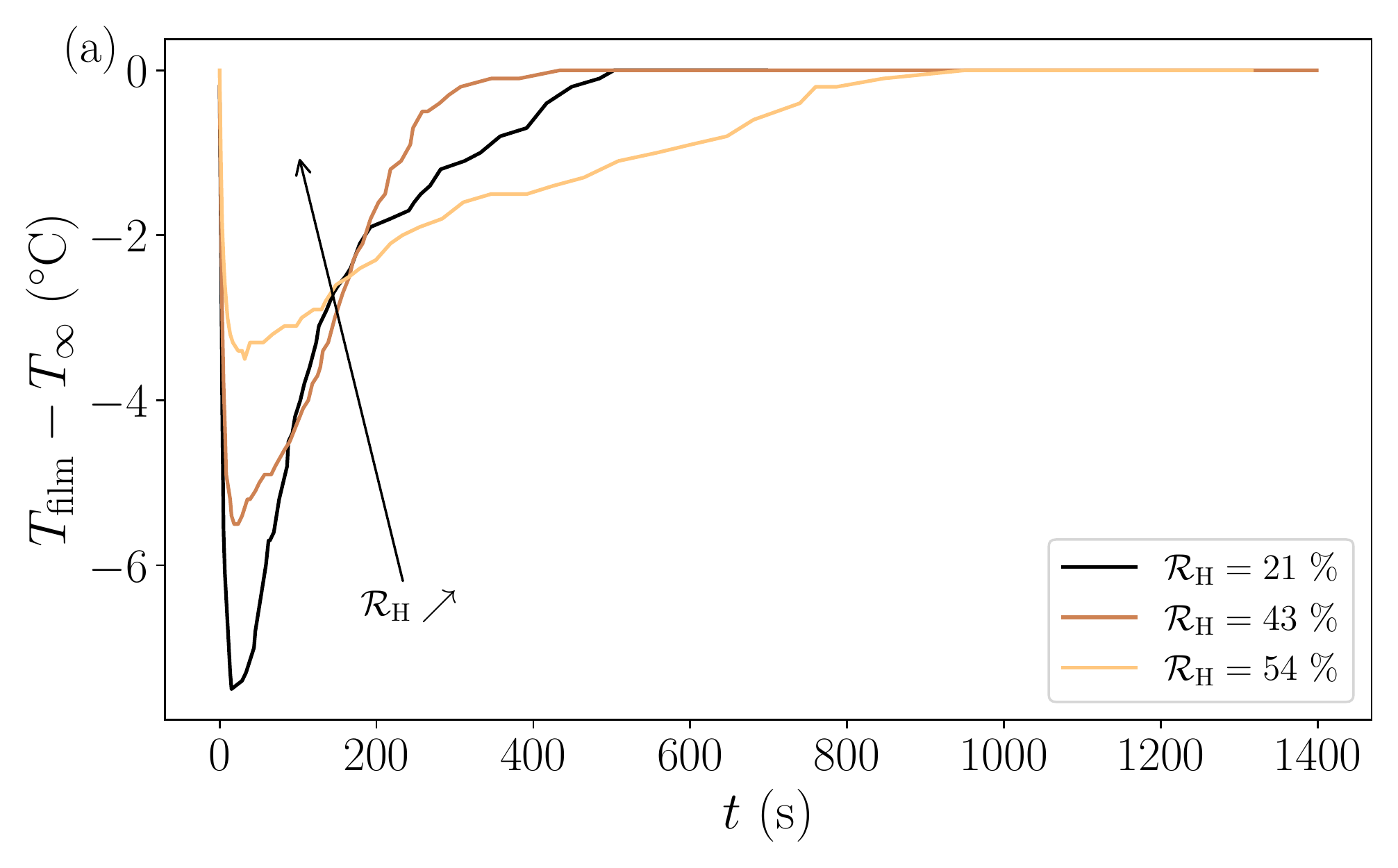}\\
    \includegraphics[width=\linewidth]{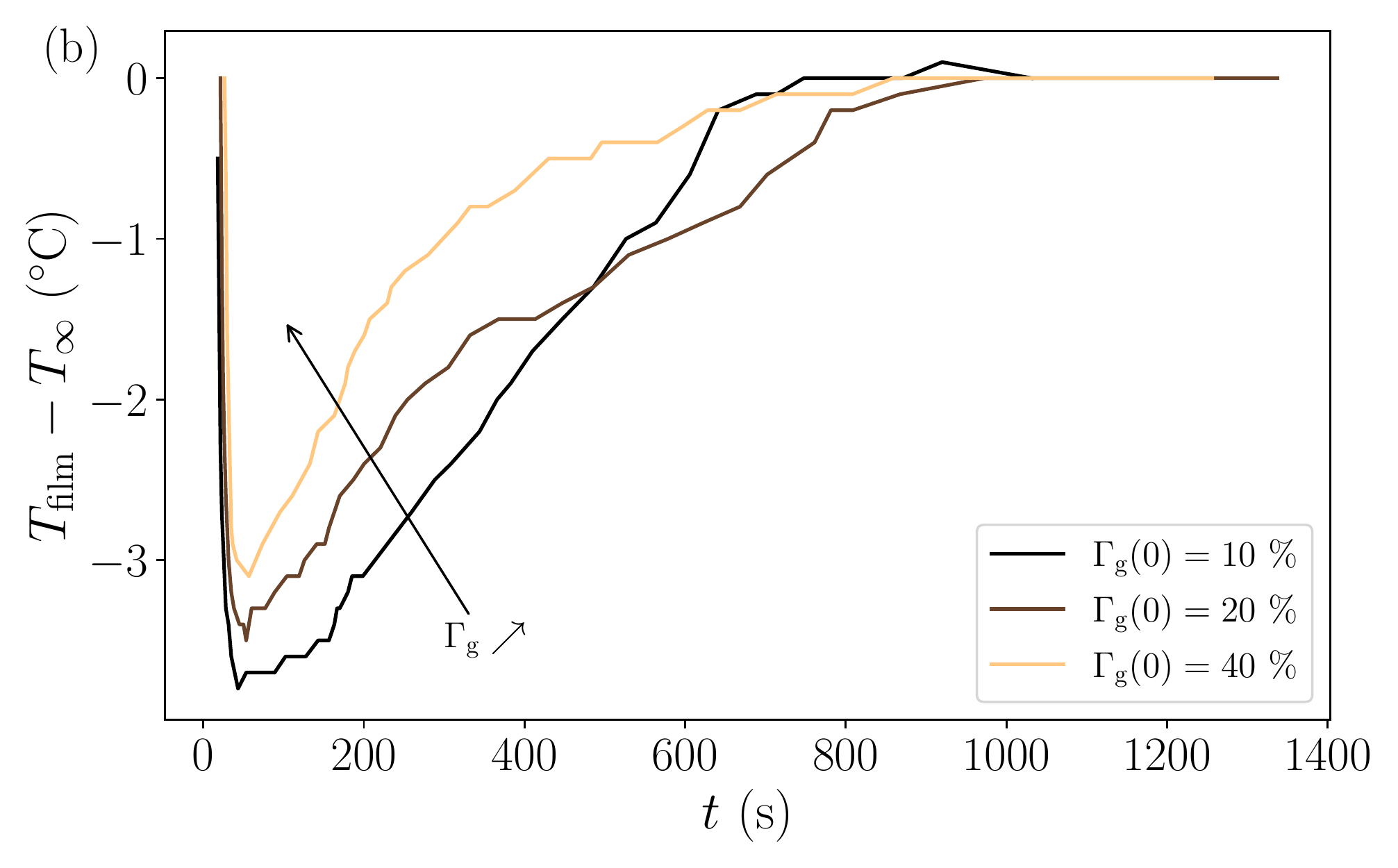}\\
    \caption{Time evolution of the temperature difference $T_{\rm film} - T_\infty$ 
    (a) for different relative humidity values ($\Gamma_{\rm g}(0)=20$~\%), and (b) for different initial concentrations of glycerol (${\cal R}_{\rm H}=54$~\%).
    The ambient temperature is about $T_\infty = 21~^\text{o}$C.
    }
    \label{fig:temperature_dynamics}
\end{figure}

To measure the soap film temperature, we use two thermocouple probes (type K, NiAl-NiCr, diameter 0.2~mm, RS PRO), both connected to a digital thermometer (RS PRO 1314).
One of the temperature probes is modified to make a ring of radius $R=6$~mm as shown in figure~\ref{fig:setup} and we check prior experiments that the measured temperatures are identical.

We performed systematic measurements of the soap film temperature in time $t$ for different humidity values ${\cal R}_{\rm H}$ and initial glycerol concentrations $\Gamma_{\rm g}(0)$, as shown in figure~\ref{fig:temperature_dynamics}.
Note that due to the manual soap film production, a delay of about 5 s may exist.
Plots presented in figure~\ref{fig:temperature_dynamics} show that the temperature difference $T_{\rm film} - T_\infty$ suddenly decreases after the soap film is produced, until a minimum value $T_{\rm film}^{\rm min}$ followed by an increase reaching a thermal equilibrium with the environment.
For an initial glycerol concentration $\Gamma_{\rm g}(0)= 20$~\%, figure~\ref{fig:temperature_dynamics}(a) indicates that the temperature drop is more pronounced at low relative humidity values, reaching $-7.5~^\text{o}$C at ${\cal R}_{\rm H}=21$~\%.
Considering a variation of the initial glycerol concentration at a constant relative humidity, the results presented in figure~\ref{fig:temperature_dynamics}(b) indicate that a larger glycerol concentration leads to a lower temperature depression.

Here, we understand that, as evaporation proceeds, the latent heat of vaporization decreases the temperature of the soap film, a cooling effect balanced by the thermal exchange between the soap film and the environment.
In addition, evaporation causes the glycerol concentration to increase in the soap film, which decreases the saturated vapor pressure and thus the evaporative flux, as observed in figure~\ref{fig:weight}.
Consequently, the slowdown of the evaporation rate reduces the cooling effect through the latent heat, such that the film temperature increases until the equilibrium is reached.
Now that a qualitative mechanism is suggested, we propose to quantify the cooling effect $\Delta T^\star = T_\infty - T_{\rm film}$ assuming a quasi-steady-state dynamics.

In addition, we assume that the film temperature is uniform.
To justify forthcoming assumptions, we consider some timescales associated to the heat and mass transfers.
We denote ${\cal D}$ the diffusion coefficient of vapor in air, $\lambda_{\rm air}$ the thermal conductivity of air, $c_{\rm p}$ the heat capacity of the liquid, and $\rho_\ell$ ($\rho_{\rm air}$) the density of the liquid (air).

The timescale to obtain a stationary regime for the mass transfer can be estimated as $R^2/{\cal D}\simeq 1$ s, for ${\cal D}= 2.5\times 10^{-5}$~m$^2$/s.
Similarly, for the temperature field in the vapor phase, the timescale is $ R^2/\alpha_{\rm air}\simeq 1$ s, for the air thermal conductivity $\alpha_{\rm air} = 19 \times10^{-6} $~m$^2$/s.
These timescales justify that at the resolution of the experiment, the evaporation and the heat flux in the atmosphere are in a steady state regime.
In addition, we will consider that the soap film temperature is uniform.
Indeed, a temperature difference between the center and the edge is rapidly damped by the thermal diffusion in the gas phase.
The timescale can be estimated as the ratio between the heat energy difference in the liquid film along the radius and the heat power in the gas phase, which writes $\rho_{\ell} c_{\rm p} Rh / \lambda_{\rm air} \simeq 1$ s, for $\lambda_{\rm air}=0.02$~W/m/K.
The timescale for heat diffusion across the film thickness $\rho_\ell c_{\rm p} h^2/\lambda_{\ell} \simeq 10^{-5}$~s being small compared to the other timescales, we can consider that the soap film temperature $T_{\rm film}$ is uniform.

In the stationary regime, the evaporative flux along the radial coordinate $r$ of the interface writes \cite{Cooke1967,Lebedev1965}
\begin{equation}
    j_{\rm ev}(r) = -\frac{2{\cal D} }{\pi R} \left[c_{\rm sat} (\Gamma_{\rm g}, T) - c_\infty(T_\infty)\right]  \left( 1 - \frac{r^2}{R^2} \right)^{-1/2},
\end{equation}
where $c$ denotes the mass concentration of vapor.
The total flux, on each interface is $Q_{\rm ev}=\int j_{\rm ev}\,{\rm d}S$, that yields

\begin{equation}\label{eq:Q_ev}
    Q_{\rm ev} = 4{\cal D} R \Delta c^\star,
\end{equation}
where $\Delta c^\star = c_\infty - c_{\rm sat} (\Gamma_{\rm g}, T) < 0$.

The vapor concentration can be related to the vapor pressure and we define the vapor pressure difference $\Delta p^\star = p_\infty - p_{\rm sat}(\Gamma_{\rm g}, T_{\rm film})$.
Denoting $M_{\rm air}$ and $M_{\rm w}$, the molar weights of dry air and water respectively, the vapor concentration difference can be related to the difference of vapor pressure as $\Delta c^\star \simeq \frac{\rho_{\rm air} M_{\rm w}}{M_{\rm air}} \frac{\Delta p^\star}{P}$, where $P$ is the atmospheric pressure.

\begin{figure}
    \centering
    \includegraphics[width=\linewidth]{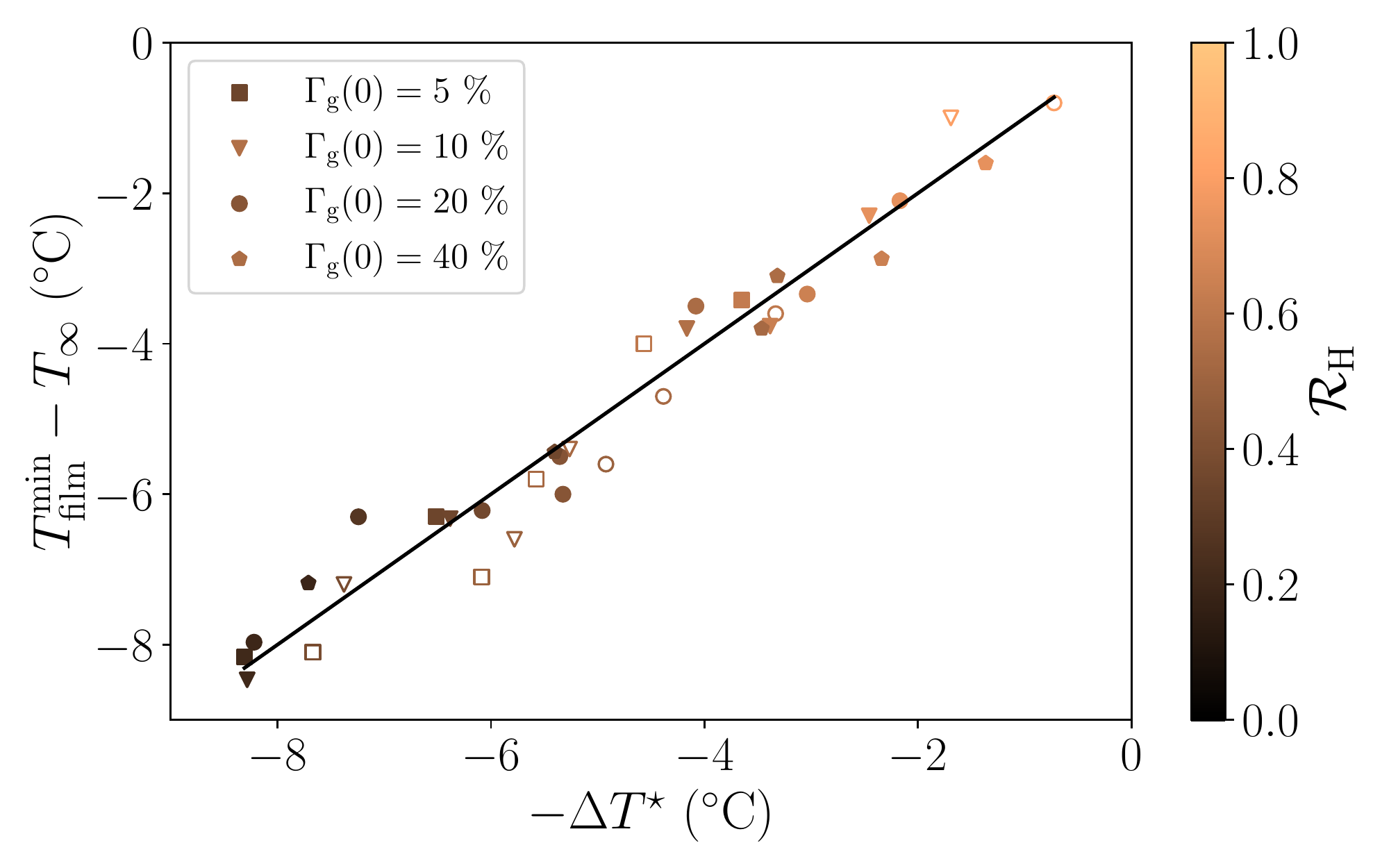}
    \caption{
    Maximum cooling effect $T_{\rm film}^{\rm min} - T_\infty$ as a function of prediction $-\Delta T^\star$ obtained with equation~\ref{eq:psychrometric} for $\Gamma_{\rm g} = \Gamma_{\rm g}(0)$.
    Open symbols are for film radius $R=2$~mm and filled symbols for $R=6$~mm.
    The relative humidity is encoded by color and the initial glycerol concentration by the symbols.
    The solid line represents equality between the two axes.
    The relative humidity values vary between 20 and 80~\%.
    }
    \label{fig:min_temperature}
\end{figure}

The temperature difference caused by evaporation leads to thermal fluxes.
We consider that the thermal exchange is mainly located at the liquid-vapor interface, neglected the role of the thin wire constituting the frame.

The thermal flux is analogous to the heat flux and writes, for each interface \cite{Cooke1967,Lebedev1965},
\begin{equation}
    Q_{\rm h} = 4 \lambda_{\rm air} R \Delta T^\star.
\end{equation}
The temperature difference also sets a radiative flux modeled by the Stefan-Boltzmann equation
\begin{equation}\label{eq:Q_rad}
    Q_{\rm rad} = \pi R^2 \epsilon \sigma (T_\infty^4 - T_{\rm film}^4),
\end{equation}
where $\sigma \simeq 5.67 \times 10^{-8}\, {\rm W} \cdot {\rm m}^{-2} \cdot {\rm K}^{-4}$ is the Stefan-Boltzmann constant and $\epsilon$ the emissivity.
The emissivity is about 0.96 for water \cite{Brewster1992}.

In the stationary regime, we write the balance of heat fluxes
\begin{equation}\label{eq:balance}
     Q_{\rm ev}\,h_{\rm ev} = -   Q_{\rm h} \left( 1 +  \frac{Q_{\rm rad}}{Q_{\rm h}}  \right).
\end{equation}
The same argument applies for the temperature profile, as the thermal diffusion coefficient is of the same order of magnitude as the vapor diffusion coefficient. 
An hypothesis that we made to establish equation~\ref{eq:balance} is to neglect the convective air flows that can arise from the natural air motion generated in particular from the film production and from the variation of air density near the film due to vapor concentration and temperature.
We can estimate that the order of magnitude of the air flow is of the order of several mm/s from the temperature gradients and the maximum radius encountered in our experiments \cite{Ostrach1953}.
Convection affects both the evaporation rate and the thermal flux, which can be written $f_{\rm ev}Q_{\rm ev}$ and $f_{\rm h}Q_{\rm h}$, respectively, where $f$ is a ventilation factor.
In gas, both ventilation factors are nearly identical and scale as the square root of the Reynolds number \cite{Ranz1952}.
For the characteristic air velocities, the ventilation factor remains close to unity and acts only in the ratio $Q_{\rm rad} / Q_{\rm h}$ in equation~\ref{eq:balance}.
Thus, we neglect air convection in this model for the small soap films involved in the experiments.

For small temperature differences $(T_\infty - T_{\rm film}) / T_\infty \ll 1 $, we have

\begin{equation}\label{eq:Q_ratio}
    \frac{Q_{\rm rad}}{Q_{\rm h}} \simeq \frac{\pi R \epsilon \sigma  T_\infty^3 }{\lambda_{\rm air}}.
\end{equation}
For $Q_{\rm rad} / Q_{\rm h}= 1$, we can define a critical radius $R_{\rm c} =\lambda_{\rm air}  / ( \pi \epsilon \sigma  T_\infty^3 )$.
At $20~^\text{o}$C, we find $R_{\rm c} = 7$~mm, which indicates that $Q_{\rm rad}$ cannot be neglected in our experiments, especially for films of 6~mm radius \cite{Hill1916}.

Substituting equations \ref{eq:Q_ev} and \ref{eq:Q_ratio} in equation \ref{eq:balance}, we obtain the so-called psychrometric equation \cite{Arnold1933} relating the vapor pressure difference $\Delta p^\star$ and the temperature difference $\Delta T^\star$
\begin{equation}\label{eq:psychrometric}
    \Delta p^\star = - P \frac{ M_{\rm air}}{\rho_{\rm air} M_{\rm w}}  \frac{\lambda_{\rm air} }{{\cal D} h_{\rm ev}} \left( 1+  \frac{\pi R \epsilon \sigma  T_\infty^3 }{\lambda_{\rm air}} \right) \Delta T^\star.
\end{equation}

Equation~\ref{eq:psychrometric} provides a prediction of the cooling effect for a given relative humidity and glycerol concentration.
As the minimum of temperature is reached during after 20 to 30 s, which leads to a variation of the glycerol concentration of 5 to 10~\% of its initial value. 
This variation modifies the vapor pressure of the solution (Eq.~\ref{eq:psatGly}) up to $2.5$~\% for $\Gamma_{\rm g}(0) = 40$~\%.
Thus, we assume that the soap film composition is weakly changed, such that we consider the initial glycerol concentration $\Gamma_{\rm g}(0)$.
In figure~\ref{fig:min_temperature}, we plot the maximum drop of temperature as a function of the temperature difference predicted by equation \ref{eq:psychrometric}, for two soap film radii (2 and 6~mm), for different relative humidity values and different initial glycerol concentrations.
Although the soap film lifetime at an initial glycerol concentration of 5~\% is not sufficient to record the full dynamics, we are able to measure the minimum temperature so that we added the data.
According to the conditions, cooling effects of $-1$ to $-8~^\text{o}$C have been measured.
These experimental measurements are in good agreement with the model, with regards to the approximations we made.

In conclusion, we have evidenced that the temperature of soap films is not necessarily equal to the ambient temperature and that the temperature difference can be significant.
The soap film studied in this paper are composed of a mixture of volatile and nonvolatile solvents, the latter compound preventing a premature rupture of the film and providing a means to tune the evaporation rate complementary to the humidity.

Experimentally, we observed that the temperature first decreases and then increases until the ambient temperature is reached again.
We reported that the magnitude of the cooling effect depends on both the relative humidity and the initial glycerol concentration, decreasing the values of these two parameters leading to stronger effects.
The cooling effect is explained by the soap film evaporation through the latent heat of vaporization.
We modeled satisfactorily the maximum cooling effect by considering a heat balance constituted of the latent heat of vaporization, the thermal conductivity from the surrounding atmosphere to the film, and the radiative flux.
This model represents a first approach of the problem and more sophisticated theoretical developments, including in particular the unsteady state, the role of natural convection for large-scale films and forced convection, will be the subject of future work.

Although soap films and bubbles have been the subject of a wide range of studies, the temperature of these objects is often considered to be equal to the environmental temperature, which is not always exact as we have demonstrated in this Letter.
In particular, the cooling effect is expected to have an influence on the fluid properties such as the viscosity and the surface tension, the surfactant critical micellar concentration.
For instance,  the variation of the  dynamic viscosity of pure water from 12~$^\text{o}$C to 20~$^\text{o}$C is $1.03~\rm{mPa}\cdot\rm{s}$ to $1.23~\rm{mPa}\cdot\rm{s}$, a variation that is more pronounced with dissolved glycerol \cite{Glycerine1963}.
More dramatically the cooling could allow the system to be below the Krafft point, potentially leading to the formation of crystals. 
Furthermore, in certain conditions, the thermal field of soapy objects may not be uniform, leading to Marangoni flows \cite{Trittel2019}.

As a result, this study suggests considering more carefully the role of evaporation through the cooling effect in the study of soap films and bubbles, especially regarding the questions raised by the scientific community on surfactant crystallization, film drainage, marginal regeneration, and film lifetime.

\section*{Acknowledgments}
We deeply thank Marina Pasquet and Marie Corpart for fruitful discussions.
This work was supported by a grant by the French National Research Agency (ANR-30519-CE30-0002).

\bibliography{biblio}

\begin{thebibliography}{10}

\bibitem{Boucher2013}
O.~Boucher, D.~Randall, P.~Artaxo, C.~Bretherton, G.~Feingold, P.~Forster,
  V.-M. Kerminen, Y.~Kondo, H.~Liao, U.~Lohmann, P.~Rasch, S.K. Satheesh,
  S.~Sherwood, B.~Stevens, and X.Y. Zhang.
\newblock {\em Clouds and Aerosols}, book section~7, page 571–658.
\newblock Cambridge University Press, Cambridge, United Kingdom and New York,
  NY, USA, 2013.

\bibitem{Murphy1998}
D.M. Murphy, J.R. Anderson, P.K. Quinn, L.M. McInnes, F.J. Brechtel, S.M.
  Kreidenweis, A.M. Middlebrook, M.~P{\'o}sfai, D.S. Thomson, and P.R. Buseck.
\newblock Influence of sea-salt on aerosol radiative properties in the southern
  ocean marine boundary layer.
\newblock {\em Nature}, 392(6671):62, 1998.

\bibitem{Leeuw2011}
G.~De~Leeuw, E.L. Andreas, M.D. Anguelova, C.W. Fairall, E.R. Lewis, C.~O'Dowd,
  M.~Schulz, and S.E. Schwartz.
\newblock Production flux of sea spray aerosol.
\newblock {\em Reviews of Geophysics}, 49(2):2010RG000349, 2011.

\bibitem{Veron2015}
F.~Veron.
\newblock Ocean spray.
\newblock {\em Annu. Rev. Fluid Mech.}, 47(1):507--538, 2015.

\bibitem{Baylor1977}
E.R. Baylor, M.B. Baylor, D.C. Blanchard, L.D. Syzdek, and C.~Appel.
\newblock Virus transfer from surf to wind.
\newblock {\em Science}, 198(4317):575--580, 1977.

\bibitem{Blanchard1989}
D.C. Blanchard.
\newblock The ejection of drops from the sea and their enrichment with bacteria
  and other materials: a review.
\newblock {\em Estuaries}, 12(3):127--137, 1989.

\bibitem{Netz2020a}
R.~R. Netz and W.~A. Eaton.
\newblock Physics of virus transmission by speaking droplets.
\newblock {\em Proc. of the Nat. Acad. of Sci.}, 117(41):25209--25211, 2020.

\bibitem{Liger-Belair2009}
G.~Liger-Belair, C.~Cilindre, R.~D. Gougeon, M.~Lucio, I.~Gebefugi, P.~Jeandet,
  and P.~Schmitt-Kopplin.
\newblock {Unraveling different chemical fingerprints between a champagne wine
  and its aerosols}.
\newblock {\em Proc. of the Nat. Acad. of Sci.}, 106(39):16545--16549, 2009.

\bibitem{Bolore2018}
D.~Bolor{\'e} and F.~Pigeonneau.
\newblock Spatial distribution of nucleated bubbles in molten glasses
  undergoing coalescence and growth.
\newblock {\em J. Am. Ceram. Soc.}, 101(5):1892--1905, 2018.

\bibitem{Gonnermann2007}
H.M. Gonnermann and M.~Manga.
\newblock The fluid mechanics inside a volcano.
\newblock {\em Annu. Rev. Fluid Mech.}, 39:321--356, 2007.

\bibitem{Poulain2018PRL}
S.~Poulain and L.~Bourouiba.
\newblock Biosurfactants change the thinning of contaminated bubbles at
  bacteria-laden water interfaces.
\newblock {\em Physical Review Letters}, 121, 11 2018.

\bibitem{Miguet2020a}
J.~Miguet, M.~Pasquet, F.~Rouyer, Y.~Fang, and E.~Rio.
\newblock Stability of big surface bubbles: Impact of evaporation and bubble
  size.
\newblock {\em Soft Matter}, 16, 2020.

\bibitem{Mysels1959}
K.~J. Mysels.
\newblock {\em Soap films: studies of their thinning and a bibliography}.
\newblock Pergamon Press, 1959.

\bibitem{Li2012}
X.~Li, S.I. Karakashev, G.M. Evans, and P.~Stevenson.
\newblock Effect of environmental humidity on static foam stability.
\newblock {\em Langmuir}, 28:4060--4068, 3 2012.

\bibitem{Pigeonneau2012}
F.~Pigeonneau, H.~Ko{\v{c}}{\'a}rkov{\'a}, and F.~Rouyer.
\newblock Stability of vertical films of molten glass due to evaporation.
\newblock {\em Colloids and Surfaces A: Physicochemical and Engineering
  Aspects}, 408:8--16, 2012.

\bibitem{Poulain2018JFM}
S.~Poulain, E.~Villermaux, and L.~Bourouiba.
\newblock Ageing and burst of surface bubbles.
\newblock {\em Journal of Fluid Mechanics}, 851:636--671, 9 2018.

\bibitem{Champougny2018}
L.~Champougny, J.~Miguet, R.~Henaff, F.~Restagno, F.~Boulogne, and E.~Rio.
\newblock Influence of evaporation on soap film rupture.
\newblock {\em Langmuir}, 34(10):3221--3227, 2018.

\bibitem{Houghton1933}
H.~G. Houghton.
\newblock A study of the evaporation of small water drops.
\newblock {\em Journal of Applied Physics}, 4(12):419--424, 1933.

\bibitem{Erbil2012}
H.~Y. Erbil.
\newblock Evaporation of pure liquid sessile and spherical suspended drops: A
  review.
\newblock {\em Adv. Colloid Interface Sci.}, 170(1–2):67 -- 86, 2012.

\bibitem{Tran2018}
H.~V. Tran, T.A.H. Nguyen, S.~R. Biggs, and A?~V. Nguyen.
\newblock On the predictions for diffusion-driven evaporation of sessile
  droplets with interface cooling.
\newblock {\em Chemical Engineering Science}, 177:417--421, 2018.

\bibitem{Beard1971}
K.~V. Beard and H.~R. Pruppacher.
\newblock A wind tunnel investigation of the rate of evaporation of small water
  drops falling at terminal velocity in air.
\newblock {\em Journal of Atmospheric Sciences}, 28(8):1455 -- 1464, 1971.

\bibitem{Andreas1995}
E.~L. Andreas.
\newblock The temperature of evaporating sea spray droplets.
\newblock {\em J. Atmos. Sci.}, 52(7):852--862, April 1995.

\bibitem{Boulogne2019}
F.~Boulogne.
\newblock Cheap and versatile humidity regulator for environmentally controlled
  experiments.
\newblock {\em The European Physical Journal E}, 42(4):51, 2019.

\bibitem{Lide2008}
D.R. Lide.
\newblock {\em CRC Handbook of Chemistry and Physics}.
\newblock 89th edition edition, 2008.

\bibitem{Glycerine1963}
Glycerine~Producers' Association et~al.
\newblock {\em Physical properties of glycerine and its solutions}.
\newblock Glycerine Producers' Association, 1963.

\bibitem{Cooke1967}
J.~R. Cooke.
\newblock Some theoretical considerations in stomatal diffusion: A field theory
  approach.
\newblock {\em Acta Biotheoretica}, 17(3):95--124, 1967.

\bibitem{Lebedev1965}
N.~N. Lebedev.
\newblock {\em Special functions and their applications}.
\newblock Courier Corporation, 1965.

\bibitem{Brewster1992}
M.~Q. Brewster.
\newblock {\em Thermal radiative transfer and properties}.
\newblock John Wiley \& Sons, 1992.

\bibitem{Ostrach1953}
S.~Ostrach.
\newblock An analysis of laminar free-convection flow and heat transfer about a
  flat plate paralled to the direction of the generating body force.
\newblock Technical report, NASA, 1953.

\bibitem{Ranz1952}
W.~E. Ranz and W.R. Marshall.
\newblock Evaporation from drops: Part {I}.
\newblock {\em Chem Eng Prog}, 48(3):141--146, 1952.

\bibitem{Hill1916}
L.~E. Hill, O.~W. Griffith, and M.~Flack.
\newblock {V}. {T}he measurement of the rate of heat-loss at body temperature
  by convection, radiation, and evaporation.
\newblock {\em Philosophical Transactions of the Royal Society of London},
  207(335-347):183--220, 1916.

\bibitem{Arnold1933}
J.~H. Arnold.
\newblock The theory of the psychrometer. {I.} {T}he mechanism of evaporation.
\newblock {\em Physics}, 4(7):255--262, 1933.

\bibitem{Trittel2019}
T.~Trittel, K.~Harth, C.~Klopp, and R.~Stannarius.
\newblock Marangoni flow in freely suspended liquid films.
\newblock {\em Phys. Rev. Lett.}, 122:234501, Jun 2019.

\end{thebibliography}

\bibliographystyle{unsrt}

\end{document}